\documentclass[%
 reprint,
superscriptaddress,
 amsmath,amssymb,
aps,
]{revtex4-2}

\usepackage{graphicx}% Include figure files
\usepackage{dcolumn}% Align table columns on decimal point
\usepackage{bm}% bold math
\usepackage{mathrsfs}
\usepackage{float} 

\begin{document}

\preprint{APS/123-QED}

\title{Bacterial turbulence at compressible fluid interfaces}% Force line breaks with \\

\author{Yuanfeng Yin}
 \affiliation{School of Physical Science and Technology, ShanghaiTech University, Shanghai, 201210, China}%Lines break automatically or can be forced with \\
\author{Bokai Zhang}%
 \email{zbk329@swu.edu.cn}
\affiliation{School of Physical Science and Technology, Southwest University, Chongqing, 400715, China
}%

\author{H. P. Zhang}%
\affiliation{School of Physics and Astronomy, Shanghai Jiao Tong University, Shanghai, China}
\affiliation{Institute of Natural Sciences, Shanghai Jiao Tong University, Shanghai, China
}%

\author{Shuo Guo}
 \email{guoshuo@shanghaitech.edu.cn}
\affiliation{School of Physical Science and Technology, ShanghaiTech University, Shanghai, 201210, China}%

\date{\today}% It is always \today, today,

\begin{abstract}
Dense bacterial suspensions at fluid interfaces provide a natural platform to explore active turbulence in a dimensional mismatch: active units are restricted to a two-dimensional surface, while the induced flows extend into the surrounding three-dimensional liquid. Using hydrophobic \textit{Serratia marcescens} at the air-water interface, we realize interfacial bacterial turbulence as a distinct class of active turbulence. The system exhibits compressible in-plane flows, with vortex size initially increasing with the thickness of the underlying fluid and saturating near 100~$\mu$m, independent of bacterial length. This behavior contrasts sharply with bulk active turbulence, where correlation length typically scales with system size. Hydrodynamic theory, together with direct measurements of the three-dimensional flow field, shows that the coupling between interfacial and bulk flows sets the emergent length scale. Our results uncover the fundamental physics of interfacial bacterial turbulence and open new strategies for geometric control of collective active flows.
\end{abstract}

\maketitle

%\tableofcontents

Motile microorganisms can spontaneously organize into large-scale, spatiotemporal chaotic flows \cite{Wensink2012,Dunkel2013,Adama2015,Blanch2018,Lin2021,Berta2019,Peng2021,Alert2022}. Unlike high-Reynolds-number inertial turbulence, active turbulence occurs at low Reynolds numbers with energy injection and dissipation at the same scales \cite{Alert2020}. This regime is characterized by the intermittent emergence and decay of vortices with characteristic length scales, and understanding as well as controlling their pattern formation and dynamics remains a central challenge \cite{Vitelli2024,Wu2024}. In bulk fluids, various hydrodynamic models consistently indicate that self-generated active stresses reorient individual swimmers, triggering the well-known Simha–Ramaswamy orientational instability \cite{Ramaswamy2002,Saintillan2008}. By contrast, in non-bulk interfacial flow environments, a rich variety of instability mechanisms and length-selection behaviors emerge, and an expanding body of experiments continues to reveal the complexity of these dynamics \cite{Gao2015,MartinezPrat2021,Sanchez2012}.

In bulk active fluids, instability lengths are effectively scale-free and determined by the system size, allowing even macroscopic confinement to control microscale collective motion \cite{Wioland2013,Lushi2014, Beppu2021, Dogic2017, Nishiguchi2017, Nishiguchi2025}. By contrast, active fluids on substrates or confined within two passive fluids acquire a finite intrinsic length scale through frictional or hydrodynamic damping \cite{Guillamat2017, Maitra2018, Gao2017}. At fluid interfaces, a distinct effect arises: flows extend both along the interface and into the underlying fluid \cite{John1993, Prasad2010, Diego2014, Salbreux2009, Huang2021}, breaking in-plane incompressibility and generating effective attractions among contractile swimmers and repulsions among extensile ones.  These effects fundamentally modify interfacial order and the associated instability mechanisms. In particular, interface-induced compressibility not only suppresses long-wavelength instabilities but also facilitates the emergence of long-range polar or nematic order, as predicted by kinetic and active liquid crystal theories \cite{Morozov2024,Maitra2023}. Such effects strongly influence vortex formation and flow organization, offering new avenues to control self-organization, mixing, and transport \cite{Singh2016,Caciagli2020,Francisca2021}. Previous studies have mainly focused on microtubule-based active nematics \cite{Sanchez2012,MartinezPrat2021} and bacterial suspensions in free-standing films \cite{Sokolov2007,Sokolov2012}, where in-plane flows remain incompressible. Yet, the collective dynamics and vortex length-scale selection in compressible interfacial active matter remain largely unexplored.

\begin{figure}[t!]
		\includegraphics[width=\columnwidth]{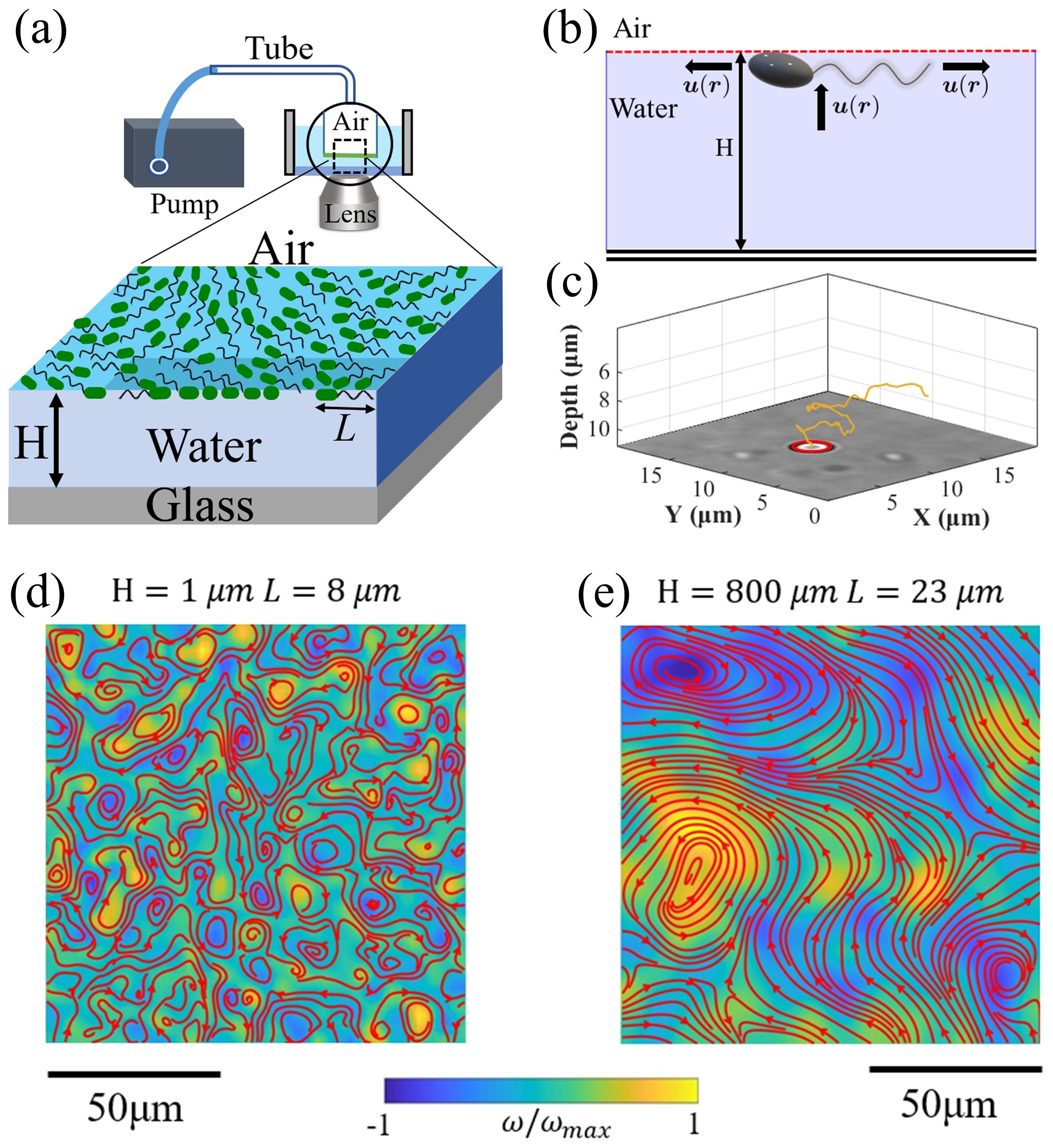}
		\caption{Experimental setup, particle tracking, and vortices at the air-water interface. 
        (a) A monolayer of hydrophobic \textit{Serratia marcescens} bacteria swims in 2D at an air-water interface. The interface is pinned and kept flat at the open end of an air-filled tube, that is partially immersed in water. H denotes the fluid thickness above the solid substrate and $L$ represents the total bacterial length, including the flagella.
        (b) Schematic illustrating a bacterium trapped at the interface.  The arrows indicate the direction of flow fields generated both at the interface and in the underlying liquid.  (c) 3D particle-tracking experiments show passive particles being advected into the interfacial layer. The red circle indicates the range of Brownian diffusion over the same time interval. 
        (d) Instantaneous flow fields, represented by vorticity color maps superimposed with streamlines, illustrate the smallest vortices (mean radius $R=3.8~\mu$m), and (e) the largest vortices ($R=53.3~\mu$m) observed under the conditions $L=8~\mu$m, H$=1~\mu$m and $L=23~\mu$m, H$=800~\mu$m.}
		\label{fig1}
\end{figure}

In this study, we realize interfacial bacterial turbulence using hydrophobic \textit{Serratia marcescens}, which are adhered to the air-water interface, forming stable monolayers atop a passive three-dimensional (3D) Stokes fluid of thickness H (Fig.~\ref{fig1}(a) and \ref{fig1}(b)). The bacterial motion is strictly two-dimensional, as confirmed by z-scan microscopy (Fig.~S1). Wild-type cells are typically $3~\mu\mathrm{m}$ long and $1~\mu\mathrm{m}$ wide, swimming at an average speed of $v_s \sim 25~\mu\mathrm{m/s}$ driven by rotating flagellar bundles. Cell length can be tuned from $1~\mu\mathrm{m}$ to $16~\mu\mathrm{m}$ by adding antibiotics at different growth stages, while cell width remains constant. Swimming speed was independent of cell length, likely due to the formation of multiple flagella~\cite{Yin22}. 

The influence of fluid thickness on swimming speed and its fluctuation is negligible (Fig.~S2(a)), suggesting that the effective viscosity induced by the underlying solid plane has a minimal impact on individual swimming behavior. When trapped by the air-water interface, bacterial trajectories exhibit greater complexity than those observed near, but not penetrating, solid–liquid or air–liquid interfaces \cite{Lauga2006, Diego2014, Leonardo2011}, displaying straight paths as well as both clockwise (CW) and counterclockwise (CCW) circular orbits (Fig.~S2(c)). Such behaviors likely result from unbalanced hydrodynamic torques and splayed flagellar configurations (Supplementary Movies 5 and 6). Over 100 trajectories were analyzed for each fluid thickness, revealing nearly equal proportions of CW and CCW swimmers (Fig.~S2(b)) and a consistent mean curvature radius of $\sim 15~\mu \mathrm{m}$, larger at air–water than at oil–water interfaces \cite{Deng2020}.

The system maintains a stable area fraction $\phi = 0.58 \pm 0.08$, ensuring fully developed active turbulence. Lower densities ($\phi < 0.4$) fail to sustain turbulence \cite{Sokolov2007}, whereas higher densities ($\phi \sim 0.72$) result in jamming. Particle-tracking experiments with passive tracer beads reveal that tracers intermittently move in and out of the interfacial plane on experimental timescales, indicating the presence of out-of-plane flow components and confirming that the surrounding fluid dynamics are 3D (Fig.~\ref{fig1}(c)). The air–water interface extends laterally ~0.75~mm, more than 20 times the largest vortex scale, minimizing boundary effects. Active turbulence is recorded in $256~\mu\mathrm{m} \times 256~\mu\mathrm{m}$ fields of view, capturing at least 10 vortices per frame. Velocity fields are extracted using particle imaging velocimetry (PIV) or optical flow analysis. The bulk fluid thickness H serves as the main control parameter in our experiments. 

Figures~\ref{fig1}(d) and~\ref{fig1}(e) display instantaneous streamlines and vorticity fields at the air–water interface for different fluid thicknesses and cell lengths. PIV analysis in dense suspensions confirmed a symmetric vorticity distribution (Fig.~S3), and the vortex size increases systematically with the both parameters. Quantitatively, the spatial velocity correlation function is defined as, $C(r)=\langle \mathbf{u}(\mathbf{x}+\mathbf{r})\cdot\mathbf{u}(\mathbf{x})\rangle/\langle \mathbf{u}^2\rangle$,
where $\mathbf{u}$ is the flow velocity and $\mathbf{r}$ the spatial separation.  The correlation is averaged over all positions and over at least 1000 frames in a fully developed turbulent state. The gradual decay of $C(r)$ reflects the localized collective motion, while its negative values at intermediate $r$ signify rotational flows (Fig.~S4). 

\begin{figure}[!htpb]
		\centering
		\includegraphics[width=0.85\columnwidth]{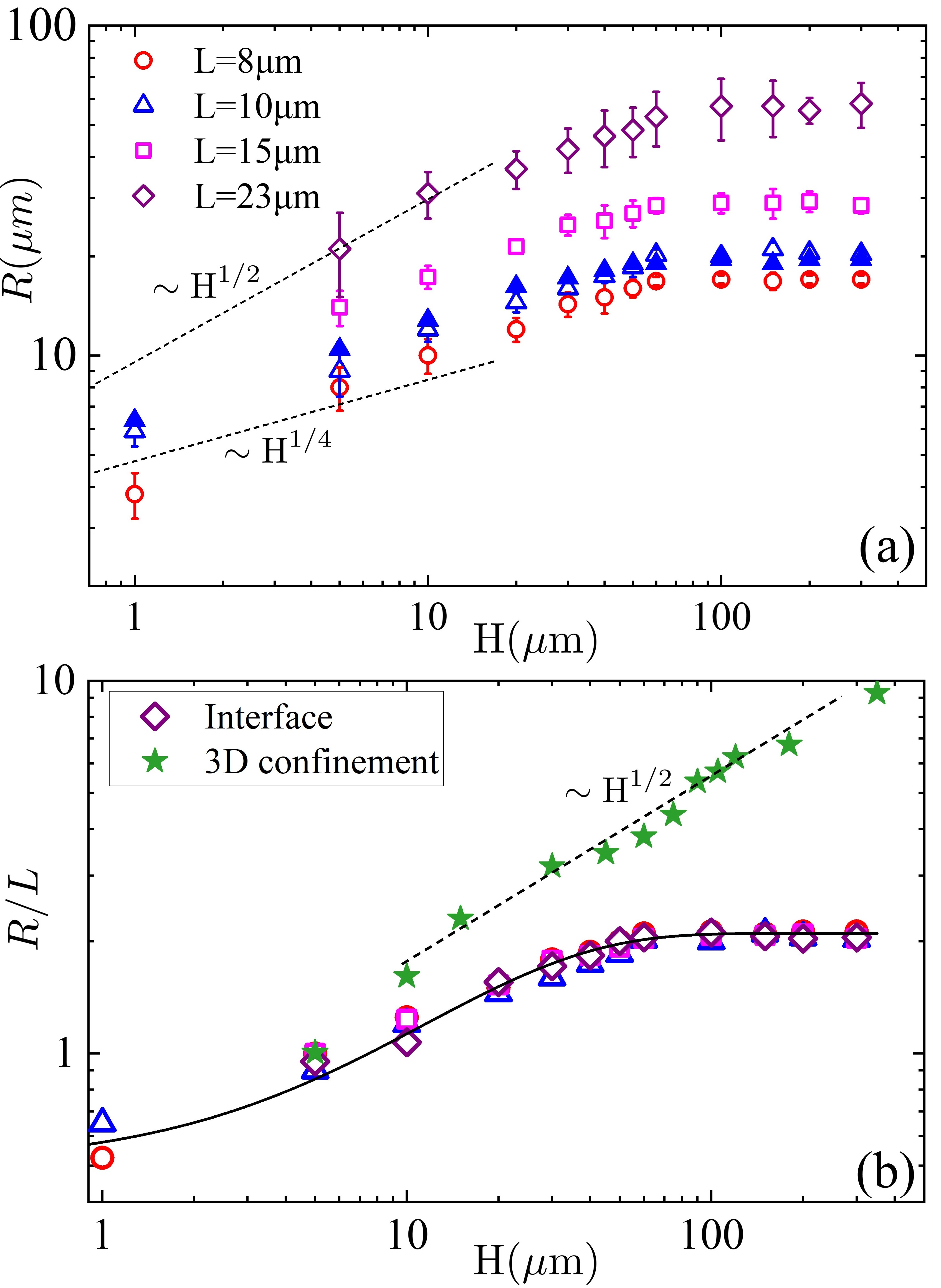}
		\caption{Fluid thickness controlling vortex size. (a) The correlation length $R$ as a function of fluid thickness H for bacteria of varying lengths $L$. Each data point represents measurements from at least three independent bacterial batches, with error bars indicating standard deviations.  Solid blue triangles denote the vortex size of wild-type bacteria ($L=10~\mu$m), calculated using the Okubo–Weiss method. 
        (b) Normalized correlation length $R/L$ for interfacial \textit{S. marcescens} suspensions (open symbols) and three-dimensional confined \textit{E. coli} suspensions (filled stars) as a function of H. The dashed line indicates an exponential fit, $R/L=2.1-1.6e^{-\text{H}/20}$, and serves as a guide to the eye.}
		\label{fig2}
\end{figure}

The characteristic vortex size is quantified by the correlation length $R$, defined by $C(R)=1/e$. As shown in Fig.~\ref{fig2}(a), $R$ increases with fluid thickness H and saturates at $\mathrm{H} \approx 100~\mu$m, indicating that coherent flows are enhanced in thicker layers. All correlation curves collapse when rescaled as $C(r/R)$ (Fig.~S4), and $R$ scales with the product of the collective velocity $V$ and correlation time (Fig.~S5) \cite{Li2019}. Consistent with observations in 3D active turbulence, $C(r)$ remains independent of bacterial density and swimming speed above the collective threshold \cite{Dunkel2013,Sokolov2007}.  

Independent measurements using the Okubo–Weiss method yield consistent vortex sizes, differing by only a constant factor of 1.2 (Fig.~S6) \cite{Ding2021}, likely due to vortex boundary definitions.  Additionally, the characteristic length extracted from the peak position of the kinetic energy spectrum exhibits the same dependence on fluid thickness as $R$ measured from the velocity correlation (Fig.~S7).  Given these consistency, $R$ is adopted as the characteristic vortex size, which increases with both H and bacterial length $L$, consistent with the direct observations in Fig.~\ref{fig1}.  Notably, when normalized by $L$, the dependence of $R$ on H collapses onto a single curve (Fig.~\ref{fig2}(b)).

%Figure~\ref{fig2}b reveals a striking scaling collapse: when the vortex size is normalized by the bacterial length $L$, data obtained for different fluid thicknesses H collapse onto a single master curve. This universal behavior reflects an underlying active length scale, $\ell_a = \sqrt{K_F/\zeta}$, which sets the characteristic size of flow structures in active nematics~\cite{Giomi2015,Hemingway2016}. Here, $K_F$ is the Frank elastic constant penalizes distortions in the bacterial orientation field, and $\zeta$ is the active stress driving the flow instability. For rod-like, flagellated bacteria, Onsager’s hard-rod theory predicts $K_F \propto n^2 L^4$~\cite{Straley1973}, while the active stress scales as $\zeta \propto nL$~\cite{Ramaswamy2002}. At constant volume fraction ($n \sim 1/L$), the resulting active length has a scaling $\ell_a \sim L$, implying a linear relationship between vortex size and bacterial length. Our experiments confirm this scaling, demonstrating that the emergent vortex size across diverse confinement thickness is governed by the intrinsic active length scale.

Several aspects of the dependence of vortex size on fluid thickness differ qualitatively from those in 3D bulk turbulence. Two fundamental distinctions between interfacial and bulk turbulence emerge. (i) In 3D bulk systems, the vortex size increases continuously with the fluid thickness H without saturation, as reported in previous studies \cite{Breuer2021, Guo2018,Peng2024,Clement2025} and confirmed by our measurements of \textit{E. coli.} using the same setup and measured at the air-water interface (Fig.~2(b)), indicating that coherent flows are system-size dependent. In contrast, in interfacial system where bacterial motion is strictly two-dimensional (2D), the vortex size saturates at a finite value that is much smaller than the lateral system size.  (ii) In confined 3D geometries, the scaling $R \sim \mathrm{H}^{1/2}$ arises from coupling between in-plane and out-of-plane orientational modes. The solid boundary imposes a standing-wave condition along the confined direction, $k_\perp = \pi / \mathrm{H}$, which selects the longest unstable wavelength \cite{Zhou2014,Dogic2020,Peng2024}. Because bacterial out-of-plane motion is suppressed at the interface, this coupling between orientational modes is broken, resulting in a weaker scaling exponent and the disappearance of vertical wavevector–mediated length-scale selection.

We employ mean-field kinetic theory (MFKT) to elucidate the qualitative distinctions in length-scale selection between interfacial and bulk (2D/3D) bacterial turbulence. In particular, we compare a purely 2D (P2D) system and an interfacial 2D (I2D) system. In both cases, the active units are confined to a plane. In the P2D system, the induced flow field is also strictly 2D, resulting in incompressible flow ($\nabla_{2D}\cdot \bm{u}_\parallel = 0$). In contrast, in the I2D system, the active units are confined to the interface but generate a 3D flow field, leading to in-plane compressibility ($\nabla_{2D}\cdot \bm{u}_\parallel \neq 0$).

MFKT models bacteria as force dipoles and construct an evolution equations for the probability distribution functions of their positions and orientations \cite{Saintillan2008,Subramanian2009,Morozov2020}. Within this framework, the homogeneous isotropic state becomes unstable once the bacterial concentration exceeds a critical value \cite{Subramanian2009}. While the governing equations for the probability density are formally identical in the P2D and I2D systems (see Eq.~\ref{eqn:5}), the resulting dynamics differ because the flow fields, entering through Eq.~\ref{eqn:6} and \ref{eqn:7}, determine the translational and rotational motion of bacteria.

\begin{figure}[!t]
	\centering
	\includegraphics[width=1.0\columnwidth]{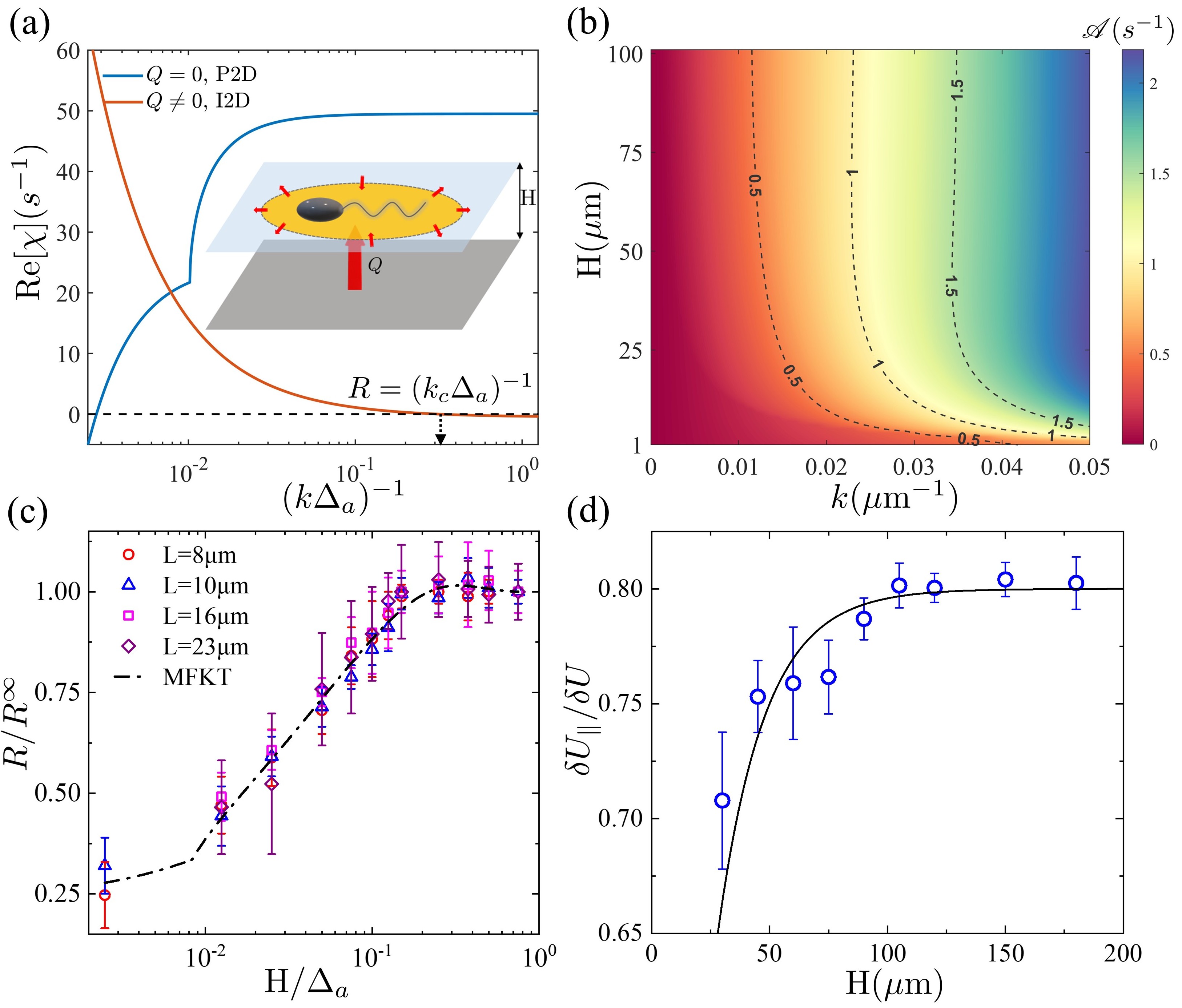}
	\caption{Growth rate, correlation lengths, viscous couplings. (a) Growth rate as a function of wavevector for the interfacial 2D system (I2D) ($ Q\neq 0$) and the purely two-dimensional system (P2D) ($ Q=0$). Inset: Schematic of the flow flux generated by a bacterium at a fluid interface. 
    (b) Contour plot illustrating the variation of $\mathscr{A}$ with respect to fluid thickness H (y-axis) and wavevector $k$ (x-axis), with the color bar indicating $\mathscr{A}$ values.  Dashed black lines represent the onset of instability, $\mathscr{A}=\lambda$,  with contour levels at 0.5, 1.0, and 1.5 $s^{-1}$.  
    (c) Normalized vortex size from experiments plotted against the instability length predicted by mean-field kinetic theory at a tumbling rate of $\lambda=0.5~s^{-1}$.  (d) Ratio of parallel to total flow velocity as a function of fluid thickness. The curve shows the same dependence on H as $R$. The dashed line represents the same exponential function as in Fig.~\ref{fig2}, scaled by a factor of $(2.1 - 1.6 e^{-\text{H}/20})/0.38$.}
	\label{fig3}
\end{figure}

In particular, MFKT predicts that the P2D system exhibits a long-wavelength instability, where the growth rate of orientational fluctuations is largest for modes with the longest wavelengths ($k^{-1}\to\infty$), as illustrated in Fig.~\ref{fig3}(a). Such long-wavelength–dominated behavior is reminiscent of 3D bulk active fluids, where the emergent length scale is governed by the largest unstable modes and ultimately truncated by the system size \cite{Saintillan2008}.

%\begin{equation}
%\chi_{P2D} = -\lambda + G\Big(\frac{v_s k}{B\kappa_{2D} %n}\Big)B\kappa_{2D} n
%\label{eqn:1}
%\end{equation}
%where $B$ is the Bretherton constant, a shape factor that takes $B=0$ for spherical objects and $B=1$ for needle-like slender rods \cite{Kim2005}. $\lambda$ is the tumbling rate, and the dipole strength is defined as $\kappa_{2D} \equiv FL/\eta_{2D}$, where $\eta_{2D}$ is the  two-dimensional viscosity. Here, $n$ represents the number density of cells, and $F$ the dipolar force generated by bacterial propulsion. The real part of the growth rate, $\mathrm{Re}[\chi_{P2D}]$, 

 %Microscopically, this reflects changes in the flux of interfacial fluid molecules~\cite{Bleibel2014,Bleibel2015,Bleibel2017}; macroscopically, it modifies the effective friction coefficient~\cite{Guillamat2016}. As a result, the scaling of fluid thickness deviates from $\text{H}^{1/2}$, exhibiting a reduced exponents of approximately 1/4 as shown in Fig.~\ref{fig2}a.

In the I2D system, where active units are confined to an interface above an underlying fluid, hydrodynamic interactions are approximated by including the first two image dipoles of strength $\kappa$~\cite{Peng2024,Mathijssen2016}. This representation captures the leading-order coupling to the bulk fluid and naturally introduces in-plane compressibility, in contrast to other dipole-based models of interfacial active layers~\cite{Gao2017, MartinezPrat2021}, which assume an incompressible in-plane flow field.

Linear stability analysis yields the interfacial growth rate as a function of the in-plane wavevector (see SM):
\begin{equation}
\chi_{I2D} = -\lambda + G\Big(\frac{2v_s f(k\mathrm{H})}{B\kappa n}\Big)\frac{B\kappa n k}{2f(k\mathrm{H})}
\label{eqn:1}
\end{equation}
Here, $\lambda$ is the tumbling rate, and $B$ is the Bretherton constant quantifying the shape factor ($B=0$ for spheres and $B=1$ for slender rods \cite{Kim2005}). The dipole strength is given by $\kappa\equiv FL/\eta_0$, where $F$ is the propulsion force, and $\eta_0$ the viscosity of water; $n$ denotes the interfacial number density. The explicit form of the function $G$ is provided in Appendix~F.

In this interfacial system, the active strength sets an intrinsic velocity scale $v_0 \equiv B\kappa n$, which is further modulated by the fluid thickness through the dimensionless geometric factor $f(x)=(2-(1-x)e^{-2x})^{-1}$ (Fig.~S12).  Further, the theory introduces an intrinsic length scale, $\Delta_a ={B\kappa n}/{\lambda}$, referred to as the \textit{active reorientation length}, which characterizes the competition between active-driven forcing and random reorientation due to tumbling.

In contrast to the P2D case, the I2D dispersion relation, Eq.~\ref{eqn:1},  exhibits the opposite dependence on wavenumber: the real part of the growth rate, $\mathrm{Re}[\chi_{I2D}]$, increases with decreasing wavelength (increasing $k$), as shown in Fig.~\ref{fig3}(a).  This trend indicates that the most unstable modes occur at short wavelengths.  In practice, however, $k^{-1}$ cannot be arbitrarily small, as it is bounded by the average interparticle spacing; below this scale the continuum approximation underlying the kinetic theory breaks down~\cite{Morozov2024}.  The absence of a long-wavelength instability therefore implies that interfacial active turbulence possesses a finite, intrinsic length scale that is independent of the system size \cite{Morozov2024}.

The opposite wavelength selection in the P2D and I2D systems can be manifested through the effective hydrodynamic interactions generated by in-plane compressibility at the interface.  To quantify this effect, we compute the hydrodynamic flux, $Q = \oint_{r=X} d\mathbf{r}\cdot \mathbf{u}$, the line integral of the in-plane velocity around a circular loop of radius $X$ centered on an active dipole. In P2D systems, incompressibility enforces $Q=0$. By contrast, for a dipole located at an interface a height H above a solid boundary (inset of Fig.~\ref{fig3}(a)),
\begin{equation}
Q = \frac{\kappa}{8X}\Big[1 - \frac{X^4 - 10 X^2 \mathrm{H}^2 + 64 \mathrm{H}^4}{(X^2 + 4\mathrm{H}^2)^{7/2}} X^3\Big] \,,
\end{equation}
which is nonzero for pushers ($\kappa>0$) due to fluid inflow beneath the interface, leading to an effective hydrodynamic repulsion. As $\mathrm{H} \to 0$, $Q \to 0$, the system recovers the P2D limit, where long-wavelength modes dominate, illustrating how interfacial compressibility modifies the wavevector dependence and wavelength selection.

To characterize the onset of orientational instability, we define the bare growth rate as $\mathscr{A}\equiv \text{Re}[G(\frac{2v_sf(k_c\text{H})}{v_0})\frac{v_0 k_c}{2f(k_c\text{H})}]$, which represents the intrinsic growth rate of orientational fluctuations in the absence of tumbling ($\lambda=0$), when each bacterium swims along a straight path.  The onset of orientational instability occurs when the real part of the growth rate changes sign, $\text{Re}[\chi_{I2D}]=0$.  According to Eq. \ref{eqn:1}, this criterion is equivalent to $\mathscr{A}=\lambda$, showing that the tumbling rate 
$\lambda$ thus counteracts the growth of orientational modes, effectively stabilizing the homogeneous isotropic state \cite{Subramanian2009}.

Figure~\ref{fig3}(b) shows the calculated $\mathscr{A}(k,\mathrm{H})$  using representative bacterial parameters: $v_s=25~\mu \text{m}/\mathrm{s}$, $v_0=200~\mu \text{m}/\mathrm{s}$, $B=1$, $\kappa=1000~\mu \text{m}^3/\mathrm{s}$, and $n=0.2~\mu \text{m}^{-2}$. The dashed lines mark the onset of instability, $\mathscr{A} = \lambda$.  Larger values of $\mathscr{A}$ indicate faster-growing orientational unstable mode.  As $k$ increases, $\mathscr{A}$ shifts to higher values, implying that shorter-wavelength modes become unstable first.  For a given $\lambda$, increasing the layer thickness H shifts the instability toward smaller $k$, corresponding to larger coherent vortices.  This trend is consistent with the experimentally observed increase of $R$ with H (Fig.~\ref{fig2}). The qualitative behavior of the instability is robust over the experimentally relevant parameter space. 

We define the characteristic vortex size as $R\equiv 2\pi/k_c$. For typical flagellated bacteria, run durations are approximately exponentially distributed with a mean of 1–2~s \cite{Beer2019}; we thus adopt $\lambda = 0.5~\mathrm{s}^{-1}$. Theoretical predictions for the normalized characteristic length, $R(\mathrm{H})/R^{\infty}$ with $R^{\infty}\equiv R(\mathrm{H}\to\infty)$, 
show excellent agreement with experimental measurements (Fig.~\ref{fig3}(c)) without adjustable parameters.  At intermediate thicknesses, the model yields a scaling $R \sim \text{H}^{1/4}$, independent of tumbling rate $\lambda$, while the scaling exponent $\alpha$ depends only weakly on the active velocity scale $v_0$ (Fig.~S8). 
At small H the system enters an oscillatory branch of the growth rate (Fig.~S11), where $\text{Im}[\chi_{I2D}]>0$, which slows the variation of $R$ with H.  In the strongly confined limit H$\to0$, expanding the function $f(x)$ gives $R\approx 3\text{H}+G(2v_s/v_0)\Delta_a/2$, indicating that the vortex size is controlled by the larger of the geometric thickness and the intrinsic active reorientation length. 

While previous theoretical studies demonstrated length selection at fluid–fluid interfaces \cite{Gao2017, Alert2020}, these analyses focused primarily on incompressible interfaces, where the selected vortex size is controlled by the viscosity of the underlying bulk fluid \cite{MartinezPrat2021}. In contrast, in our system the interfacial bacterial speed shows little dependence on H, suggesting that the confinement effect cannot be explained by the effective viscosity alone.  This leads us to propose that scale selection arises from the compressibility of the interface itself, mediated by coupling and exchange between the interfacial fluid layer and the underlying 3D bulk. Understanding this coupling is therefore key to revealing how confinement controls collective dynamics in interfacial bacterial suspensions.

To directly probe the influence of fluid thickness on 3D flows, we measured local flow fields 15~$\mu$m below the wild-type bacterial layer using optical tweezers and tracer particles, across a range of H.  The time-averaged horizontal and total flow speeds are quantified by the standard deviation of the respective velocity components, $\delta U_{||}=({\delta U_x}^2+{\delta U_y}^2)^{1/2}$ and $\delta U=({\delta U_x}^2+{\delta U_y}^2+{\delta U_z}^2)^{1/2}$. As shown in Fig.~\ref{fig3}(d), the ratio of the horizontal to total flow, denoted $\delta U_\parallel/\delta U$, exhibits an exponential dependence on the parameter H, approaching a value of 0.8 at large H. This suggests that while the horizontal flow dominates, the vertical flux remains non-negligible, underscoring the importance of 3D flow structures in the interfacial bacterial fluid systems. 

Velocity fluctuations exhibit the same H-dependence as the characteristic vortex size $R$, up to a constant normalization factor.  This behavior can be understood by considering how active stresses are dissipated: energy is predominantly balanced by 2D shear viscosity, $\eta_\mathrm{2D} k^2 \delta U_\parallel \approx \zeta \bm{k}\cdot \bm{n}\bm{n}$, while nematic order grows exponentially, $\bm{n}\bm{n}\sim exp(\chi t)$.  At the characteristic scale $k_{\text{max}}^{-1}$ selected by the maximal growth rate, horizontal flow speeds scale as $\delta U_\parallel \sim k_\mathrm{max}^{-1}$, establishing a direct connection between the measured 3D flow and the vortex size $R$.  This reveals, for the first time, how fluid confinement modulates collective interfacial dynamics through the interplay of horizontal and vertical flows, providing a mechanistic understanding of thickness-controlled vortex dynamics.

In summary, we have realized bacterial interfacial turbulence, where motile bacteria confined to an air–water interface collectively drive chaotic flows. Unlike bulk active fluids, which are generally scale-free, the interfacial bacterial layer exhibits an intrinsic length scale. Subphase fluid thickness does not affect the motility of individual bacteria, but it controls the vertical hydrodynamic coupling to the underlying flow, thereby modulating effective interactions within the interfacial layer. Within the MFKT framework, we identify a distinct mechanism of length selection at the interface, corresponding to a transition from long-wavelength instabilities in bulk active fluids to short-wavelength instabilities confined to 2D interfaces in a 3D Stokes flow. These results provide a unified picture linking the experimentally observed vortex size to hydrodynamic mechanisms, demonstrating how confinement and vertical coupling select a characteristic length scale and thus drive the transition from scale-free bulk turbulence to length-selected interfacial turbulence.

The mechanism that selects a finite intrinsic length scale in our system differs fundamentally from that in active layers on solid substrates. In the latter, substrate friction acts as a momentum sink, introducing a finite dissipation length that competes with the bulk long-wavelength instability and thereby determines a friction-controlled vortex size \cite{Guillamat2017,Maitra2018}.  In our interfacial system, the vortex size $R$ scales linearly with the bacterial body length $L$ (Fig.~\ref{fig2}(b)), indicating that the relevant cutoff is not imposed by external dissipation but arises from an intrinsic microscopic scale of the suspension.  A consistent physical picture emerges: in dense bacterial suspensions, the vortex size $R$ scales with the size of locally nematic domains. Owing to excluded-volume interactions of rod-like bacteria, this spacing is expected to be largest along the nematic director and to increase with the bacterial length $L$.  The observed collapse of $R/L$ at fixed fluid thickness H across populations with different cell lengths suggests that the emergent vortex size is proportional to the average spacing between cells. This provides a natural explanation for the linear scaling of $R$ with $L$ in interfacial bacterial turbulence. This behavior contrasts sharply with bulk active turbulence, where the characteristic vortex size shows no comparable collapse with swimmer length \cite{Peng2024}.
 
% Here, the selection mechanism of a finite intrinsic length scale differs fundamentally from that in systems directly in contact with substrates, which acts as a momentum sink, in which a lengthscale about friction leads to the finite size vortex size.  In 动量守恒的界面上，我们展示了生长率的依赖关系由于界面的可压缩性在定性上被改变了，从而截断在小尺度下，which 连续性假设所能满足的最小尺度。值得注意的是，实验上，涡旋尺寸正比于细菌长度，（fig.2b）。以下的物理图像猜想出细菌长度是与the average interparticle spacing成正比的，从而合理化涡旋尺寸是决定于the average interparticle spacing的。

%The characteristic length scale of the bacterial suspension is typically proportional to the size of locally nematic domains. Within such a domain, the maximum interparticle distance occurs along the nematic director and scales with the bacterial length $L$. As shown in Fig. \ref{fig2}(b), for a fixed H, the ratio $R/L$ remains constant across bacterial populations of different lengths. These results suggest that, in interfacial turbulence, the interparticle spacing scales proportionally with the emergent characteristic length of the system.

Interfacial active turbulence is expected to emerge under various geometric constraints, such as vesicle or droplet surfaces \cite{Saintillan2025}, and to facilitate nontrivial surface-associated transport phenomena, including nutrient mixing on surfaces and the transport of biomolecules within membranes embedded in bulk fluids~\cite{Deng2024,Arratia2024, Manikantan2020,Rafael2021}.  In 2D active interfaces with conserved particle number, effective hydrodynamic interactions and vortical flow are tunable, offering a versatile platform for observing and manipulating diverse self-organizing phenomena and collective motions, such as active transport \cite{Hu2021,Mathijssen2018, Robert2024}, defects dynamics \cite{Head2024, Aranson2017,Head2024a}, and hyperuniformity \cite{Voigt2024}.

\textit{Acknowledgments}—We thank Xiaozhou He, Xiaqing Shi, Fanlong Meng and Yongxiang Huang for useful discussion. This work is supported by the startup support of ShanghaiTech University (S. G.), the National Natural Science Foundation of China under grant No. 32071253 (S. G.), the Fundamental Research Funds for the Central Universities under Grant No. SWU-KT25031 (B. Z.), and the National Natural Science Foundation of China under grant No. 12374218 and 12347102 (B. Z.), No. 12225410 and No. 12074243 (HPZ), and the National Key R\&D Program of China under Grant no. 2021YFA0910700 (HPZ). 

\bibliography{ref}% Produces the bibliography via BibTeX.
\clearpage

\section*{End Matter}
\textit{Appendix A: Cell culture}—A single colony of wild-type \textit{S. marcescens} bacteria  (strain ATCC 274) is transferred into 2 ml of liquid Terrific Broth and incubated overnight (14–16 h) at 37$^\text{o}$C in a shaker at 250 rpm. The culture is diluted 100-fold into another 3 ml of Terrific Broth and cultured at 30$^\text{o}$C in the same shaker. To obtain bacteria with a normal length of $L=10~ \mu \text{m}$, the diluted suspension is cultured for 5 h. To obtain shorter bacteria with $L=8~ \mu $m, antibiotic A22 (5 $\mu$g/mL) is added into the dilution before the second step culture. To obtain filamentous bacteria with a body length $L= 16~ \mu $m, the diluted suspension is first cultured for 3 h. Then, antibiotic Ampicillin, Sodium Salt (100 $\mu$g/mL) and Tween20 (0.1\%) are added, and the culture is incubated for an additional 2 h. To obtain bacteria with $L=23~ \mu $m, the diluted suspension is first cultured for 1.5 h. Ampicillin and Tween20 of the same concentration are then added followed by another 3.5 h culture. The length of time after the addition of Ampicillin determines the final length of the bacteria. The cell culture is transferred into centrifuge tubes and washed twice with deionized water using a centrifuge at 1200 g. Finally, the cell culture is concentrated and ready to use. The width and the hydrophobicity of the cells and their swimming speed are found to be independent of the cell length. The density of bacteria in a partially-occupied interface (area fraction of the cell bodies), denoted as $\phi$, is determined by the total number of bacteria in a dispersion, cell shape and the interface area, can be adjusted by the initial bulk concentration. 

\vspace{1em}
\textit{Appendix B: Sample preparation and control of external fluid thickness}—
We developed a microfluidic device to control the thickness of external fluid beneath the 2D bacterial layer.  After washing with mobility buffer, $100~\mu L$  suspension of swimming \textit{S. marcescens} with bulk concentration $\sim 1.6\times 10^9/mL$ is placed into an open liquid cell with a diameter of $1~cm$ and a depth of $1~mm$, which has a glass substrate for observation under a microscope (Nikon Ti-2), as shown in Fig.~\ref{fig1}(a).  Instead of using the existing air-water interface in the open cell, we create a new air-water interface in an air-filled glass tube with an inner diameter of $0.75~mm$, that is partially immersed into the bacterial suspension. The tube is set perpendicular to the glass substrate, as shown in Fig.~\ref{fig1}(a). The inner and outer surface of the tube is treated with trichloro(1H, 1H, 2H, 2H-perfluorooctyl) silane (FTS). By connecting the other end of the tube (inlet) to a gas pressure controller (FluidicLab PC1) and increasing the pressure, the new 'internal' air-water interface is pinned at the open end of the tube without any slipping and is kept flat with a radius of curvature $>2~ mm$. The height of the internal interface relative to the substrate can be adjusted by vertically moving the tube using a micro-manipulator (Narishige MMO-4, accuracy $0.1~\mu $m). After the formation of the internal interface, bacteria accumulate at this interface and the surface density becomes stable after 5 minutes. The large liquid reservoir in the liquid cell ensures the stability of the new interface over a wide range ($1~\mu \text{m}<\text{H}<1000~\mu$m) with minimal evaporation effects. This stability is advantageous for the study of active matter at interfaces. For heights of a millimeter or larger, a liquid cell without the microfluidic device can be used directly. Active turbulence is recorded using a camera (Photometric Prime) at a frame rate of approximately 40 frames per second.

%\begin{figure}[!t]
%	\centering
%	\includegraphics[width=\columnwidth]{fig4.jpg}
%	\caption{Ratio of parallel to total flow velocity as a function of liquid thickness. The curve shows the same dependence on $H$ as $R$. The dashed line represents the same exponential function as in Fig.~\ref{fig2}, scaled by a factor of $(2.1 - 1.6 e^{-H/20})/0.38$.}
%	\label{fig4}
%\end{figure}

\vspace{1em}
\textit{Appendix C: Flow velocity beneath the liquid surface}—We measure the flow velocity in the liquid beneath the 2D bacterial turbulence using silica spheres of diameter $3~\mu m$, held by an optical tweezer (Aresis Tweez305). The local flow velocity $\bf{U}$ is determined from the low-Reynolds-number force balance $k\bf{\Delta X}=\xi (\bf{U}-\bf{\Delta \dot{X}})$, where $\bf{\Delta X}$ is the displacement of the trapped particle relative to the tweezer center, and $\xi=6\pi \eta a$ is the viscous drag coefficient. The horizontal position of the particle is obtained by standard particle tracking, while its vertical position is obtained according to the brightness and quality factor of the particle image \cite{Peng2024}. The temporal fluctuations of the flow velocity exhibits zero mean in all three directions and follow Gaussian distributions.  The velocity fluctuations in three directions, $\delta U_x$, $\delta U_y$, and $\delta U_z$, are measured independently three times (Fig.~S9).

\vspace{1em}
\textit{Appendix D: Smoluchowski equation}—A Doi-Onsager kinetic theory is developed to describe the orientational instability of bacteria swimming at the interface and to extract the corresponding vortex size.  The mean-field kinetic theory has been widely applied to 3D and 2D dilute bacterial suspensions \cite{Saintillan2008}.  Recently, \v{S}kult\'{e}ty \textit{et al}. extended this theory to 2D sheets embedded in 3D fluids \cite{Morozov2024}, uncovering distinct short-wavelength instabilities that differ significantly from the long-wavelength instabilities observed in bulk 2D and 3D fluids.  Here, we further generalize the approach of \v{S}kult\'{e}ty \textit{et al.} to the air-water interface, incorporating the effects of in-plane compressibility and variable fluid thickness.

We consider a suspension of $N$ bacteria swimming at an interface of area $S$. The system is described by a one-particle density distribution function, $\Psi(\bm{x},\bm{p},t)$, which gives the probability density of finding a bacterium at position $\bm{x}$ with orientation $\bm{p}$ at time $t$.  The temporal evolution of $\Psi$ follows the Smoluchowski equation derived from particle-number conservation~\cite{Subramanian2009},

\begin{equation}
\frac{\partial}{\partial t} \Psi + \frac{\partial}{\partial x^\alpha} \left( \dot{x}^\alpha \Psi \right) + \mathbb{P}^{\alpha \beta} \frac{\partial}{\partial p^\beta} \left( \dot{p}^\alpha \Psi \right) = -\lambda \Psi + \frac{\lambda}{2 \pi} \int d\bm{p} \, \Psi
\label{eqn:5}
\end{equation}
where the Greek letter $\alpha$ denotes the 2D Cartesian components $x$ and $y$.  $\mathbb{P}^{\alpha\beta}=\delta^{\alpha\beta}-p^{\alpha}p^{\beta}$ is the projection operator onto the unit $\bm{p}$ sphere, and $\delta^{\alpha\beta}$ is the Kronecker delta function.  The right-hand side describes the random tumbling events, with $\lambda$ being the tumbling rate.  The probability density function is normalized as 
$\int d\bm{x}d\bm{p}\Psi(\bm{x},\bm{p},t)=1$. The translational and rotational motion of a single swimmer are given by
\begin{align}
\dot{x}^\alpha &= v_s p^\alpha + \mathcal{U}^\alpha(\bm{x})\label{eqn:6},\\
\dot{p}^\alpha &= \left( \delta^{\alpha \beta} - p^\alpha p^\beta \right) \left( \mathcal{W}^{\beta \gamma}(\bm{x}) + B \mathcal{E}^{\beta \gamma}(\bm{x}) \right) p^\gamma
\label{eqn:7}
\end{align}
where $v_s$ is the swimming velocity, and $\mathcal{U}^{\alpha}=N\int d\bm{x}' d\bm{p}' u^{\alpha}(\bm{x}-\bm{x}',\bm{p}')\Psi(\bm{x}',\bm{p}',t)$ is the fluid velocity induced by all other bacteria.  Eq.  \ref{eqn:7} is the well-known Jeffery's orbit, the reorientation of a bacterium due to the local velocity gradient. The tensor   $\mathcal{W}^{\beta\gamma}=\frac{1}{2}(\partial^{\beta}\mathcal{U}^\gamma-\partial^\gamma \mathcal{U}^\beta)$ and $\mathcal{E}^{\beta\gamma}=\frac{1}{2}(\partial^\beta \mathcal{U}^\gamma+\partial^\gamma \mathcal{U}^\beta)$ represent the vorticity and rate-of-strain tensors, respectively.  The Bretherton parameter $B=(A^2-1)/(A^2+1)$ characterizes the cell shape in terms of the aspect ratio $A$. 

\vspace{1em}
\textit{Appendix E: Linear stability analysis}—In the isotropic and homogeneous state, the steady distribution is $\Psi_0 =\frac{1}{2 \pi S}$.  A perturbation around this base state is introduced as, $\Psi(\bm{x},\bm{p},t)=\frac{1}{2\pi S}+\delta \Psi(\bm{x},\bm{p},t)$.  Linearizing the Smoluchowski equation with respect to $\delta \Psi$ yields, in Fourier space,
\begin{equation}
[\partial_t+\lambda+iv_s \bm{p}\cdot\bm{k}]\delta \hat{\Psi}=\frac{\lambda}{2\pi}\delta\hat{\rho}+\frac{n}{2\pi}[2Bp^{\alpha}p^{\beta}-(1+B)\delta^{\alpha\beta}]ik^{\alpha}\delta\hat{\mathcal{U}}^{\beta}
\label{eqn:8}
\end{equation}
where $n=N/S$ is the bacterial area density. The density and velocity fluctuations are defined as
\begin{align}
\delta \hat{\rho}(\bm{k},t)&=\int d\bm{p} \delta \hat{\Psi}(\bm{k},\bm{p},t)\label{eqn:9},\\
\delta\hat{\mathcal{U}}^{\alpha}(\bm{k},t)&=\int d\bm{p}\hat{u}^{\alpha}(\bm{k},\bm{p})\delta\hat{\Psi}(\bm{k},\bm{p},t).\label{eqn:10}
\end{align}
Assuming $\delta \hat{\Psi}(\bm{k},\bm{p},t)=\delta\hat{\Psi}(\bm{k},\bm{p})e^{\chi t}$, the real part of $\chi$ is the growth rate and determine the linear stability of small perturbations to the isotropic and homogeneous state.  Substituting this ansatz into Eq. (\ref{eqn:8}) yields a closed set of equations for $\delta \hat{\rho}$ and $\delta \hat{\mathcal{U}}^\alpha$:
\begin{align}
&\delta \hat{\rho} = i n \delta \hat{\mathcal{U}}^{\alpha}\int \frac{d\bm{p}}{2\pi}\Big[\mathcal{J}^\alpha(\bm{p},\bm{k},\chi,B)+\delta \hat{\rho}\frac{\lambda}{\mathcal{L}(\chi,\bm{p}\cdot \bm{k})}\Big]\label{eqn:11}\\ 
&\delta\hat{\mathcal{U}}^\alpha =in\delta\hat{\mathcal{U}}^\beta\int\frac{d\bm{p}}{2\pi}\Big[\hat{u}^\alpha(\bm{k},\bm{p})\mathcal{J}^\beta(\bm{p},\bm{k},\chi,B)+\delta\hat{\rho}\frac{\lambda \hat{u}^\alpha(\bm{k},\bm{p})}{\mathcal{L}(\chi,\bm{p}\cdot \bm{k})}\Big]
\label{eqn:12}
\end{align}
where $\mathcal{J}^\alpha=(2Bp^\alpha\bm{p}\cdot\bm{k}-(1+B)k^\alpha)/(\chi+\lambda+iv_s\bm{p}\cdot\bm{k})$ and $\mathcal{L}(\chi,\bm{p}\cdot\bm{k})=\chi+\lambda+iv_s\bm{p}\cdot\bm{k}$. Without loss of generality, we take the wavevector $\bm{k}$ along the $x$-axis. The longitudinal and transverse components of the velocity fluctuation are then $\delta\mathcal{\hat{U}}_\parallel=\delta\mathcal{\hat{U}}^x$ and $\delta\mathcal{\hat{U}}_{\perp}=\delta\mathcal{\hat{U}}^y$, respectively.  The solution depends on the $k$-space velocity field at the air–water interface above a solid substrate. An approximate expression for this velocity field, obtained by considering the first two image reflections, is provided in the SM.

\textit{Appendix F: Growth rate equation}—The transverse velocity fluctuations determine the orientational instability and are decoupled from the density fluctuations.  Solving the eigenvalue problem defined by Eq.~(\ref{eqn:11}) and (\ref{eqn:12}) (see SM) yields
\begin{equation}
\chi_{I2D}=-\lambda+G\Big(\frac{2v_sf(k\text{H})}{B\kappa n}\Big)\frac{B\kappa n k}{2f(k\text{H})}
\label{eqn:14}
\end{equation}
where $G(x)=6x^2[4-(1\pm i\sqrt{3})/h(x)-(1\mp i\sqrt{3})h(x)]^{-1}$ and $h(x)=[54x^2-1+6\sqrt{3}x\sqrt{27 x^2-1}]^{1/3}$.  The influence of fluid thickness on hydrodynamic interactions is captured by the function $f(x)$,
\begin{equation}
f(x)=\frac{1}{2-(1-x)exp(-2 x)}
\label{eqn:15}
\end{equation}
The growth rate of orientational fluctuations is defined as the real part of the eigenvalue $\chi_{I2D}$, $\mathscr{A}\equiv \text{Re}[\chi_{I2D}]$. The condition $\mathscr{A}=0$ determines the onset of instability and the corresponding onset wavevector $k_c$. 

For a P2D system, the explicit expression of growth rate of orientational fluctuations is given by \cite{Morozov2024},
\begin{equation}
    \chi_{P2D} = -\lambda + G\Big(\frac{v_sk}{B\kappa n}\Big)B\kappa n
\end{equation}
Here, only the function $G$ depends on the wavevector. Since $\mathrm{Re}[G(x)]$ is a monotonically decreasing function, $\chi_{P2D}$ reaches its maximum in the long-wavelength limit, $k \to 0$.

\end{document}